\documentclass[sigconf,nonacm,10pt]{acmart}

\usepackage[normalem]{ulem}
\usepackage{colortbl}
\usepackage{xcolor}
\usepackage{graphicx}
\usepackage{caption}
\usepackage{subcaption}
\usepackage{wrapfig}

\newcommand{\coloredtext}[2]{\textcolor{#1}{#2}}

\newcommand{\name}{\textit{SoilSound}}
\newcommand{\labMAE}{5.01\%}
\newcommand{\outdoorMAE}{\coloredtext{black}{2.39\%}}
\newcommand{\unseenMAE}{\coloredtext{black}{3.65\%}}

\AtBeginDocument{%
  }

\begin{document}

\title[SoilSound]{SoilSound: Smartphone-based Soil Moisture Estimation}

\author{Yixuan Gao}
\email{yixuan@cs.cornell.edu}
\orcid{0000-0003-1778-3104}
\affiliation{%
  \institution{Cornell Tech}
  \city{New York}
  \state{New York}
  \country{USA}
}

\author{Tanvir Ahmed}
\email{tanvir@infosci.cornell.edu}
\orcid{0000-0002-9468-5033}
\affiliation{%
  \institution{Cornell Tech}
  \city{New York}
  \state{New York}
  \country{USA}
}

\author{Shuang He}
\email{sh2646@cornell.edu }
\orcid{0009-0001-4603-0315}
\affiliation{%
  \institution{Cornell Tech}
  \city{New York}
  \state{New York}
  \country{USA}
}

\author{Zhongqi Cheng}
\email{ZCheng@brooklyn.cuny.edu}
\orcid{0000-0002-2486-9662}
\affiliation{%
  \institution{Brooklyn College of the City University of New York}
  \city{New York}
  \state{New York}
  \country{USA}
}

\author{Rajalakshmi Nandakumar}
\email{rajalakshmi.nandakumar@cornell.edu}
\orcid{0000-0002-1601-148X}
\affiliation{%
  \institution{Cornell Tech}
  \city{New York}
  \state{New York}
  \country{USA}
}

\renewcommand{\shortauthors}{Gao, et al.}

\begin{abstract}
Soil moisture monitoring is essential for agriculture and environmental management, yet existing methods require either invasive probes disturbing the soil or specialized equipment, limiting access to the public.
We present \name{}, an ubiquitous accessible smartphone-based acoustic sensing system that can measure soil moisture without disturbing the soil. We leverage the built-in speaker and microphone to perform a vertical scan mechanism to accurately measure moisture without any calibration. Unlike existing work that use transmissive properties, we propose an alternate model for acoustic reflections in soil based on the surface roughness effect to enable moisture sensing without disturbing the soil.

The system works by sending acoustic chirps towards the soil and recording the reflections during a vertical scan, which are then processed and fed to a convolutional neural network for on-device soil moisture estimation with negligible computational, memory, or power overhead. 
We evaluated the system by training with curated soils in boxes in the lab and testing in the outdoor fields and show that \name{} achieves a mean absolute error (MAE) of \outdoorMAE{} across 10 different locations. Overall, the evaluation shows that \name{} can accurately track soil moisture levels ranging from 15.9\% to 34.0\% across multiple soil types, environments, and users; without requiring any calibration or disturbing the soil, enabling widespread moisture monitoring for home gardeners, urban farmers, citizen scientists, and agricultural communities in resource-limited settings.






\end{abstract}






\maketitle

\section{Introduction}


 Soil moisture is a measurement of the water held within the pore spaces among soil particles. It is influenced by factors such as weather, soil composition, and the plants growing in it.   
It is an environmental variable that governs many processes on and below the ground, and provides a snapshot of the health and condition of the landscape.
Real time and high spatial resolution soil moisture data is essential for maintaining healthy ecosystems and supporting agricultural productivity. It is of particular importance for precision agriculture.

For in situ soil moisture measurements, probe-based sensors  are the most common because of their portability and accuracy \cite{rao2011moisture, ojo2015calibration, metergroup_teros12}, while research-grade sensors can often cost \$250 to \$10,000 USD \cite{kodikara2014soil}. However, these methods are  invasive - disturbing the soil in the process of measurement by probe insertion. Existing non-invasive methods such as Ground Penetrating Radar (GPR) requires bulky specialized equipment \cite{steelman2011comparison}. Recent methods, based on radio frequency (RF) sensing \cite{ding2019towards, feng2022lte, wang2020soil, jiao2023soiltag, ding2023soil}, or acoustic sensing \cite{adamo2004acoustic, oelze2002measurement, meisami2013site, song2024regional, voldan2024moisture}, depend on specialized hardware such as software-define radio (SDR), or dedicated embedded systems with custom antenna/microphone/speakers. While smartphone camera-based solutions offer accessibility by analyzing soil color changes, their accuracies are very limited since soil color depends heavily on organic matter content and mineral composition \cite{taneja2022predicting, sakti2018estimating}. 


In this paper, we propose \name{}, a smartphone-based wireless acoustic sensing system to measure soil moisture without disturbing the soil. \name{} is based on the observation that reflection of acoustic signals by the soil is highly correlated with soil moisture levels. We leverage the built-in speaker-microphone pair, where the speaker sends a frequency-modulated continuous wave (FMCW) acoustic signal and the microphone subsequently records the echo from soil which is then processed to estimate soil moisture. 
However, there are two key challenges when designing a smartphone-friendly, non-soil-disturbing acoustic sensing system for moisture measurements. 

First, prior work on acoustic-based soil moisture estimation have mostly relied on the \textit{transmissive} nature and property of acoustic signals in soil, which require speakers and microphones buried at different positions in the soil-- not only disturbing the soil but also inapplicable for existing smartphone hardware \cite{adamo2004acoustic, oelze2002measurement, meisami2013site, song2024regional}. For moisture detection, the \textit{reflective} nature (i.e., how soil reflect acoustic signals) is still under-explored. Towards this, we propose an alternate model based on the surface roughness effect: increase in moisture content in soil reduces surface roughness, that in turn increase specular reflection of signal by the soil. This alternate model enables us to quantify soil moisture using reflected acoustic signals without disturbing the soil. 
The second challenge, most commercial off-the-shelf (COTS) smartphone devices have limited number of speaker and microphones, and thus are often insufficient for array processing algorithms. 
We solve this problem by proposing a novel \textit{vertical} scan approach, where the user places the smartphone close to the soil
surface and slowly move it upward at a constant speed while the device transmits and receives FMCW acoustic chirps signal. By analyzing the reflected signals  across multiple observation heights, we can distinguish genuine soil reflections from multi-path noise and environmental
interference. Since smartphone microphone has a finite aperture, this approach also allows recording direct reflections from a larger surface area around the normal axis.

Our system \name{} achieves an accuracy comparable to that of standard soil moisture sensors and is generalizable to different soil types. The users neither have to calibrate the system nor dig the soil, put it in a box, or place any probe or tags inside the soil. \name{} generates FMCW acoustic chirp signal (linearly sweeping from 7 to 22 kHz) and records the soil reflections at multiple heights, preprocesses the signal and uses a convolutional neural network to estimate soil moisture. We built \name{} as an android smartphone app that uses 2.2 MB memory and can infer moisture under one second ($745 \pm 89 $ms).

We evaluate \name{} through extensive experiments validating its performance: accuracy, generalizability, and usability. We first perform controlled lab experiments with prepared soil samples in separate boxes, consisting of different soil types with varying soil moisture contents, ranging from 10\% to 38\% volumetric water content (VWC). We achieve \labMAE{} MAE when trained and tested on the same soil type, and \unseenMAE{} MAE when tested on different soil types. We also conduct a study with five users to show that our system is usable beyond the lab. We then test our system in the field without disturbing the soil and achieve an MAE of \outdoorMAE{} across 10 different locations with variable field settings, The purpose is also to assess the system's resilience to ambient acoustic noise and environmental variables. 

In summary, we make the following contributions:
\begin{itemize}
  \item We build the first acoustic-based wireless soil moisture estimation system for commodity smartphones which is easy-to-use, accessible, and works without any probes, tags, or disturbing the soil. A handheld vertical scanning mechanism using the smartphone application provides soil moisture estimation under a second completely running on-device.
  
  \item We propose and validate experimentally an alternate model for acoustic reflections in soil, revealing that increased moisture reduces soil scattering and enhances specular reflection, providing accurate moisture estimation without soil disturbance or soil-specific calibration.
  
  
  \item We validate \name{} with both in-lab and field experiments across multiple soil types, demonstrating cross-domain generalization to unseen soil types, and confirming usability through user studies with untrained participants.
\end{itemize}

\section{Related Work}

\begin{table}
    \caption{Comparison of soil moisture sensing approaches.}
    \label{tab:moisture_sensing}
    \centering
    \scriptsize
    \begin{tabular}{|l|l|l|}
    \hline
    \textbf{System} & \textbf{Hardware} & \textbf{Error}\\
    \hline 
    \hline
    \multicolumn{3}{|c|}{\textbf{Methods that disturb the soil for moisture estimation}} \\
    \hline
    TDR/FDR/Capacitance \cite{rao2011moisture, ojo2015calibration, metergroup_teros12} & Handheld sensor \& probes & 1-4\% \\
    \hline
    Neutron Probes \cite{kodikara2014soil} & Handheld sensor \& probes & 1-4\% \\
    \hline
    Acoustic \cite{adamo2004acoustic, oelze2002measurement, meisami2013site, song2024regional, voldan2024moisture} & Custom hardware & 0.1-2\% \\ 
    \hline
    Strobe~\cite{ding2019towards} & USRP \& 3 antennas (in soil) & $\leq$3\% \\
    \hline
    LTE-Soil~\cite{feng2022lte} & SDR \& 2 antennas (in soil) & 3.15\% \\
    \hline
    GreenTag~\cite{wang2020soil} & RFID reader \& tags (in soil) & 5\% \\
    \hline
    SoilTAG~\cite{jiao2023soiltag} & RFID reader \& tags (in soil) & 2-8\% \\
    \hline
    SoilId~\cite{ding2023soil} & UWB on drone \& reflectors & 0.05\% \\
    \hline
    \hline
    \multicolumn{3}{|c|}{\textbf{Methods that do not disturb the soil for moisture estimation}} \\
    \hline
    CoMEt~\cite{khan2022estimating} & USRP & 1.1\% \\
    \hline
    GPR \cite{steelman2011comparison} & Heavy sensor on pushcarts & 1-3\% \\
    \hline
    Satellite (low spatial resolution) \cite{entekhabi2010soil} & L-Band radar \& Radiometer & 4\% \\ 
    \hline
    \rowcolor[HTML]{C7EDCC}
    \textbf{SoilSound [Ours]} & \textbf{Smartphone only} & \textbf{\outdoorMAE{}} \\
    \hline
    \end{tabular}
\end{table}

Soil moisture measurement has evolved from manual gravimetric sampling to diverse electronic sensing modalities. While traditional contact methods achieve high accuracy, the need for non-invasive, scalable solutions has driven research toward wireless sensing approaches. In this section we examine the different techniques to measure soil moisture. At a high level, we classify the existing systems into three categories: contact-based, RF, and acoustic sensing systems. Table \ref{tab:moisture_sensing} provides a comparison of these systems with \name{}.

\subsection{Contact-based Methods}
The gold standard for soil moisture measurement remains direct contact methods that exploit the dramatic change in soil's dielectric properties with water content, from $\varepsilon_r \approx 3$-$5$ (dry) to $\varepsilon_r \approx 20$-$30$ (wet) \cite{cihlar1974dielectric}. Time Domain Reflectometry (TDR) measures electromagnetic pulse propagation through soil via inserted probes, achieving 2-3\% accuracy based on Topp's equation relating permittivity to moisture~\cite{topp1980electromagnetic, ledieu1986method}. Frequency Domain Reflectometry (FDR) and capacitance sensors operate on similar principles at lower cost (\$40-250) but with reduced accuracy and increased sensitivity to soil salinity and temperature, as a result, requiring field calibration \cite{ojo2015calibration, cominelli2024calibration}. Despite their accuracy, these methods share fundamental limitations: they require physical probe insertion (10-20 cm), disturb soil structure, provide only point measurements, and remain impractical for rapid large-area surveys. 

Other well-known contact-based soil moisture sensors include resistive sensors (measuring change of resistance in a porous material), neutron probes (detecting density of emitted neutrons from radioactive sources), and thermal probes (measuring thermal conductivity and diffusivity). Although expensive ($\sim\$10,000$), manual, and time consuming (~3 minutes per location), neutron probes are considered the most accurate method for measuring the volumetric water content of soil \cite{kodikara2014soil}. On the other hand, resistive and thermal methods are inexpensive, but suffer in accuracy, long-term stability, higher power consumption, and overall generalizability, which make them unsuitable for reasearch-grade measurements \cite{rasheed2022soil, jorapur2015low}. These constraints have motivated the development of non-contact wireless approaches that trade some accuracy for dramatically improved coverage and convenience.

\subsection{RF wireless Sensing Methods}
The transition from contact to wireless sensing leverages electromagnetic or acoustic wave interactions with soil moisture, primarily through resulting changes in dielectric permittivity or the velocity. We classify them into two categories - system that require burying antennas or system that can working without disturbing the soil.

\subsubsection{Soil-disturbing}
Close-range RF systems have demonstrated the feasibility of non-contact measurement by adapting existing wireless infrastructure. Strobe~\cite{ding2019towards} repurposes WiFi signals, analyzing phase and amplitude shifts across multiple antennas to estimate soil permittivity despite WiFi's limited bandwidth. This approach achieves 5-7\% accuracy but requires precise antenna positioning and complex inverse problem solutions to separate moisture from other soil properties, disturbing the soil in the process.

Building on RF principles, SoilId~\cite{ding2023soil} employs unmanned aerial vehicle (UAV)-mounted Ultra-Wideband (UWB) radar, leveraging UWB's larger bandwidth for improved resolution. By analyzing reflections between the soil surface and a buried corner reflector, it achieves depth-specific measurements. However, the requirement for pre-buried reflectors limits practical deployment. Other systems explore alternative RF bands: LoRa-based sensors decode phase shifts for low-power operation~\cite{kiv2022smol, chang2022sensor}, RFID tags offer passive sensing through impedance changes~\cite{wang2020soil, jiao2023soiltag}, ground radar coupled with back-scatter tags buried in soil \cite{josephson2021low}. These approaches reduce infrastructure requirements but still need specialized hardware or in-soil placement, thereby requiring disturbing the soil. Besides RF, there are some acoustic-based soil moisture sensing algorithms in the existing literature, but they still suffer from the soil-disturbing nature. We review acoustic based methods separately in the next section as it is the key for this research. Fully non-disturbing, accurate, and accessible soil moisture sensing is still an under-explored.

\subsubsection{Non-disturbing}
At very large scales (9-36 km spatial resolution), satellite remote sensing provides regional moisture monitoring through microwave radiometry or synthetic aperture radar. Primarily designed to distinguish
frozen from thawed land surfaces for agricultural planning and climate studies, these methods measure only the top 2-5 cm of soil with kilometer-scale spatial resolution \cite{entekhabi2010soil}. Ground Penetrating Radar (GPR) bridges the gap between point sensors and satellite coverage, offering depth profiles with centimeter resolution, but requires equipment costing thousands of dollars and expertise in data interpretation \cite{steelman2011comparison}. CoMEt \cite{khan2022estimating} is a portable soil moisture sensing solution, but complex hardware requirement is still limiting its accessibility. Lastly, some camera-based methods have been explored in this area \cite{taneja2022predicting, sakti2018estimating}, but they are heavily influenced by the soil organic and mineral matter, which makes them unreliable for everyday use.

\subsection{Acoustic Sensing Approaches}
While RF approaches dominate wireless soil moisture sensing in the recent researches, acoustic methods offer complementary advantages: lower hardware cost, reduced power consumption, and immunity to electromagnetic interference. Several works have showed that the soil moisture can be mapped by accurately by determining the velocity of sound inside the soil \cite{adamo2004acoustic, oelze2002measurement, meisami2013site, song2024regional}. However, current algorithms rely on transmitting sound wave inside a box and record it on the other end, requiring not only to disturb the soil but also use high power audio amplifiers. The reflective properties of low-power acoustic signals have not been explored well, which is essential for implementing in commodity devices. Acoustic sensing has proven effective for material characterization in industrial settings, with ultrasonic sensors (>40 kHz) measuring thickness, detecting flaws, and even estimating moisture in wood products~\cite{voldan2024moisture}. At the other extreme, seismic methods (<100 Hz) characterize large-scale geological structures, including groundwater aquifers.

The intermediate frequency range (1-22 kHz) accessible to consumer audio hardware remains surprisingly underexplored for material detection. 
%
This presents an opportunity: smartphones are ubiquitous, their acoustic hardware continues improving, and this frequency range offers reasonable propagation in air while maintaining sufficient bandwidth for ranging. Our work demonstrates that this overlooked frequency band, combined with appropriate signal processing and modeling, enables practical soil moisture measurement using only commodity smartphone hardware—eliminating the cost and deployment barriers of existing wireless approaches.
\section{Modeling Soil Moisture from Acoustic Signal Reflections}
Our goal is to detect soil moisture without disturbing the soil using a COTS device. This requires a reflective approach, where acoustic signals are sent and received from the same side of the air-soil interface. However, modeling how soil reflects acoustic signals is challenging.

\subsection{Challenge: Soil moisture does not affect acoustic reflection coefficient}

\subsubsection{Reflection coefficient for soil}

Consider an acoustic signal transmitted by a speaker and reflected by the soil. 
If we assume the air-soil interface as a boundary between two different media, the reflection coefficient is determined by the impedance mismatch between them. According to classical acoustic theory, the power reflection coefficient at a planar interface is given by:

\begin{equation}
    R_0 = \left( \frac{Z_{\text{soil}} - Z_{\text{air}}}{Z_{\text{soil}} + Z_{\text{air}}} \right)^2 = \left( \frac{\rho_{\text{soil}}v_{\text{soil}} - \rho_{\text{air}}v_{\text{air}}}{\rho_{\text{soil}}v_{\text{soil}} + \rho_{\text{air}}v_{\text{air}}} \right)^2
\end{equation}

where the acoustic impedance $Z = \rho v$ is the product of the density of the medium $\rho$ and the velocity of the sound $v$.

In standard conditions, the acoustic impedance for air is well-established at $Z_{\text{air}} = 415 \text{ Rayls}$ ($\rho_{\text{air}} = 1.21 \text{ kg/m}^3$, $v_{\text{air}} = 343 \text{ m/s}$). However, soil properties vary significantly with moisture content. Measurements by Oelze et al. \cite{oelze2002measurement} show that dry agricultural soil exhibits a density around $1300 \text{ kg/m}^3$ and a strong velocity near $450 \text{ m/s}$, yielding $Z_{\text{dry}} \approx 5.85 \times 10^5 \text{ Rayls}$. As water infiltrates the pores, both density and velocity increase dramatically. Saturated soil reaches densities of approximately $2000 \text{ kg/m}^3$ with velocities approaching $1500 \text{ m/s}$, resulting in $Z_{\text{sat}} \approx 3.0 \times 10^6 \text{ Rayls}$.

Despite this high increase in the acoustic impedance from dry to saturated conditions, the reflection coefficient remains nearly constant, because $Z_{\text{soil}} \gg Z_{\text{air}}$ in both cases. Substituting these values, we find $R_0 = 0.9972$ for dry soil and $R_0 = 0.9994$ for saturated soil, a negligible increase of only $0.22\%$. This analysis suggests that the acoustic reflections from the soil surface should be consistently strong and virtually independent of the moisture content if only the reflection coefficient is considered.

\subsubsection{Experimental observation}
To test the reflection coefficient theory described above, we implemented an acoustic FMCW (frequency modulated continuous wave) sensing system using a smartphone speaker and a microphone. 
The system transmits a chirp signal whose frequency linearly increases from $f_0$ to $f_1$ within a time interval $T$. The signal s(t) is given by

\begin{equation}
s(t) = A \cdot \cos\left(2\pi \left(f_0 t + \frac{B}{2T} t^2 \right)\right)
\end{equation} 

where $B = f_1 - f_0$ is the sweep bandwidth. When this signal reflects off objects at different distances, the echoes return with different time delays. By mixing the received signal with the original transmitted signal, these time delays are converted into distinct beat frequencies, creating a \textbf{range profile} that shows the reflection strength at each distance. The range resolution of the sensing system is given by $\Delta R = v/(2B)$, where $v$ is the speed of sound. For a chirp signal with a bandwidth of 15~KHz, we can achieve range resolution of $\approx 1.1~cm$.

\begin{figure}[hbtp]
    \vspace{-10pt} 
    \centering
    \includegraphics[width=\linewidth]{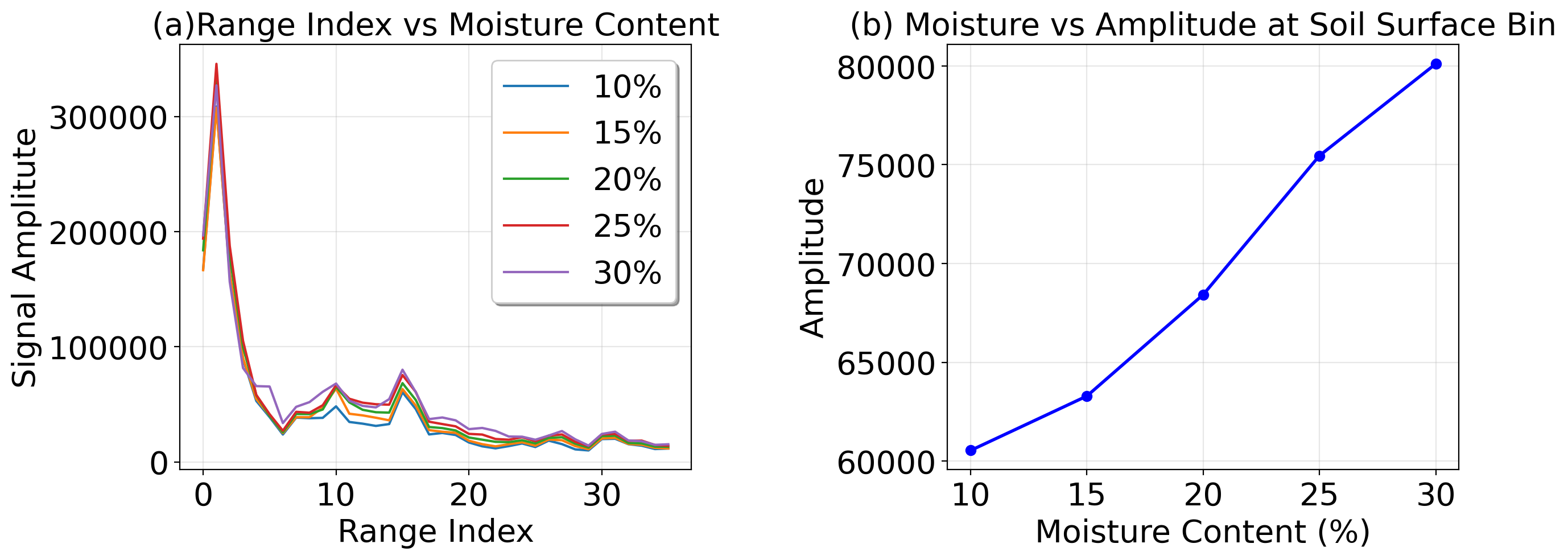}

    \caption{Experiment of the FMCW acoustic system mounted at a fixed location: (a) range profiles at various moisture levels, (b) reflection amplitude at the soil surface range bin.}
    \vspace{-10pt} 
    \label{fig:wet_vs_dry_echo}
\end{figure}

In our experiment, we performed the above FMCW sensing with soil samples of varying moisture levels . As shown in Figure~\ref{fig:wet_vs_dry_echo}, the acoustic measurements reveal a contradiction to the predictions based on reflection coefficient alone, that the power of the direct reflection increases with increase in moisture level. Figure~\ref{fig:wet_vs_dry_echo}(a) shows the range profiles for different moisture levels, where the peak around range bin 15 corresponds to direct reflection from the surface of the soil (the shortest acoustic path). Figure~\ref{fig:wet_vs_dry_echo}(b) demonstrates that the echo amplitude in this bin increases with the moisture content from dry to wet conditions. This trend cannot be explained by the impedance mismatch alone, suggesting the presence of more complex underlying physical mechanisms.

\subsection{Our Approach}

Based on above experiment, we note that the roughness of the soil surface is another critical factor that influences the reflection behavior. The key insight is that as more water is added to the soil, the surface becomes smoother, leading to more specular reflection and thus an increased amplitude in the corresponding range bin. We can model this behavior of the soil based on existing work.

\subsubsection{Modeling Specular Reflection of Soil}

Darmon et al.~\cite{darmon2020acoustic} extended the Kirchhoff approximation~\cite{beckmann1987scattering} to acoustic signals and showed that at normal incidence, specular reflection strength $P_{\text{spec}}$ depends exponentially on surface roughness:
\begin{equation}
    P_{\text{spec}} = P_{\text{inc}} \cdot R_0 \cdot \exp[-g^2]
    \label{eq:P_vs_g}
\end{equation}
where $P_{\text{inc}}$ is the incident signal power and g is the Rayleigh roughness parameter given by:
\begin{equation}
    g = \frac{2\pi\sigma_h}{\lambda}
    \label{eq:rayleigh_parameter}
\end{equation}
with $\sigma_h$ representing surface roughness and $\lambda$ the acoustic wavelength.

Dry soil presents a granular, unconsolidated surface with roughness of 5-10 mm for sandy soils. As moisture increases, capillary forces create menisci between particles, generating inter-particle cohesion exceeding 10 kPa~\cite{lu2006suction}. This cohesion binds particles together, reducing surface roughness.

Based on principles of unsaturated soil mechanics~\cite{lu2006suction} and experimental observations~\cite{ravindran2022effect}, moisture-dependent surface roughness can be modeled as:
\begin{equation}
    \label{eq:sigma_theta}
    \sigma_h(\theta_v) = \sigma_{h,\text{dry}} \cdot \exp(-\alpha \cdot S_e^\beta) + \sigma_{h,\text{sat}}
\end{equation}

This equation describes how the surface roughness $\sigma_h$ evolves as the soil transitions from dry to wet conditions. At the core of this model is the effective saturation $S_e = (\theta_v - \theta_r)/(\theta_s - \theta_r)$, which normalizes the volumetric water content $\theta_v$ (water volume per total soil volume) to a 0-1 scale using the residual moisture $\theta_r$ that remains even in 'dry' soil due to tightly bound water films  and the saturation point $\theta_s$ where all pore spaces are filled with water. $\alpha$ is the decay rate factor describing how soil roughness decays with moisture, and $\beta$ is nonlinearity exponent for $S_e$. This normalization allows us to compare moisture effects across different soil types with varying water-holding capacities. 

The model represents the mechanism that dry soil maintains its granular, rough texture with characteristic roughness $\sigma_{h,\text{dry}}$, but as water infiltrates, capillary forces progressively bind particles together, smoothing the surface towards a minimum roughness $\sigma_{h,\text{sat}}$. The exponential term governs this transition, with parameter $\alpha$ controlling the initial rate of smoothing when water first enters the soil, and $\beta$ shaping how this smoothing progresses toward saturation. 

For instance, the typical values of $\sigma_{h,\text{dry}}$ for sandy soils is 5-10 mm reflecting the coarse grain structure, which reduces to $\sigma_{h,\text{sat}}$ = 0.5-1 mm when water films create a relatively smooth surface. The transition parameters $\alpha$ = 3-5 and $\beta$ = 1.5 reflect sand's rapid initial response to moisture due to its large pore spaces ~\cite{allmaras1966total}. If we transmit 15~Khz frequency acoustic signal (($\lambda$ = 23 mm) then based on equation 4, the Rayleigh parameter $g$ for the sandy soil decreases from approximately 1.9 when dry (strong scattering regime) to 0.4 when the soil is moist (predominantly specular regime). 

 This dramatic change in surface roughness theoretically increases the specular reflection power by 34 times. Based on this model, if our system can estimate the specular reflection then we can measure soil moisture. However, in an acoustic sensing system the microphone records both specular and diffuse components. 

\subsubsection{Modeling specular reflection and scattering}

Consider an FMCW sensing system placed at a height $h$ above the surface of the soil. The range bin corresponding to distance $h$ represents the shortest possible propagation path, a wave traveling vertically downward to the surface and directly back to the sensor. This is the only path that produces a specular (mirror-like) reflection, since specular reflections occur along the shortest, normal-incidence path.  The scattering signal will show up in every range bin, including and after the direct reflection range bin. For the sensor height of $h$, the total received reflection power$P(r)$ at each range bin $r$ can be modeled as:

\begin{equation}
    \label{eq:reflection_decomposition}
    P(r) = \begin{cases}
    P_{\text{spec}}(r) + P_{\text{diff}}(r) \cdot \Omega_{\text{eff}}(h) & \text{if } r \approx h \text{ (direct path)} \\
    P_{\text{diff}}(r) \cdot \Omega_{\text{eff}}(h) & \text{if } r > h \text{ (oblique paths)}
    \end{cases}
\end{equation}

where only a fraction $\Omega_{\text{eff}}(h)$ of diffuse scattering is captured in each range bin after r, due to the finite transducer aperture. Here, $\Omega_{\text{eff}}(h)$ is defined by the effective solid angle at height h representing the net fraction of diffusely scattered energy captured by the receiver.

It is worth noting that the extended range signals do not represent reflections from subsurface layers. The extreme impedance mismatch at the air-soil interface yields a transmission coefficient of only $T_0 = 1 - R_0 \leq 0.0028$, which means that less than 0.3\% of the acoustic power penetrates the soil. Combined with severe attenuation (50-80 dB/m at 10 kHz in moist soil~\cite{oelze2002measurement}), subsurface reflections are effectively undetectable. 

Based on this theory, we can formulate the received power at each range bin $r$ and sensor height $h$ as a function of moisture content and soil parameters. From Equations~\ref{eq:reflection_decomposition}, \ref{eq:P_vs_g}, and \ref{eq:sigma_theta}, the specular reflection can be expressed as function of $\theta_v$, the unknown moisture variable and the soil-specific parameters ($\alpha$, $\beta$, $\sigma_{h,\text{dry}}$, $\sigma_{h,\text{sat}}$, $\theta_r$, $\theta_s$). The diffuse scattering component $P_{\text{diff}}$ will depend on additional complex factors such as the height of the smartphone from the soil surface and antenna aperture. This shows that it is difficult to estimate moisture using only the range bin corresponding to the height h of the smartphone. In principle, we will need multiple independent equations, to solve for $\theta_v$, without needing to calibrate for parameters. 



Instead we identify two key insights that can be used to model the relationship between the soil moisture and the reflected signals and enable robust moisture estimation without explicit parameter fitting.

\textbf{Insight 1: Extended range bins capture scattering, providing complementary information.} 
As shown in our geometric analysis, extended range bins ($r > h$) contain only scattering reflections which is fundamentally different from combined specular and scattering signals in the direct path. By including the entire range profile, rather than relying only on the single peak at $r = h$, we gain access to these separate scattering-only measurements. These scattering patterns evolve systematically as moisture changes the soil's surface roughness, providing independent data.

\textbf{Insight 2: Measuring range profiles for multiple heights provides additional geometric perspectives.} As mentioned above, the scattering reflections recorded at the microphone depends on the height $h$ of the device from the soil.
For a receiver with aperture radius $a$, the maximum collection angle narrows with height 
as $\theta_{\mathrm{max}} \approx \arctan(a/h)$, so the effective collection solid angle 
decreases as
\begin{equation}
    \label{eq:omega_height}
    \Omega_{\mathrm{eff}}(h) \propto \left(\frac{a}{h}\right)^2, \quad (h \gg a).
\end{equation}
At lower heights when the smartphone is closer to soil, the wide collection cone captures a larger fraction of the scattered energy, 
making the measured signal more sensitive to surface roughness changes. As the height increases, 
the collection cone becomes narrower and an increasingly smaller fraction of scattered energy 
is collected, while the specular reflection at $r=h$ is always fully captured due to its 
vertical path. Thus, measurements at different heights provide different weightings between 
scattering and specular components. Since moisture affects these components differently, multi-height measurements provide complementary 
constraints that improve the robustness of moisture estimation.

\textbf{Combining both insights in a learning framework.} 
Together, these two insights form a multi-dimensional sensing strategy. Across range bins (Insight 1), we capture the specular and scattering reflections from different distances. Across sensor heights (Insight 2), we capture how the balance between reflection and scattering shifts with geometry. This produces a structured data cube where each axis encodes different physical information about the soil. Convolutional neural networks are well-suited to process such structured data, automatically learning the nonlinear mapping from acoustic signatures to moisture content. By training on diverse soil types and moisture levels, the model learns to generalize without requiring explicit parameter tuning or site-specific calibration.

\section{System Design}

\begin{figure*}[hbtp]
    \centering
    \includegraphics[width=\linewidth]{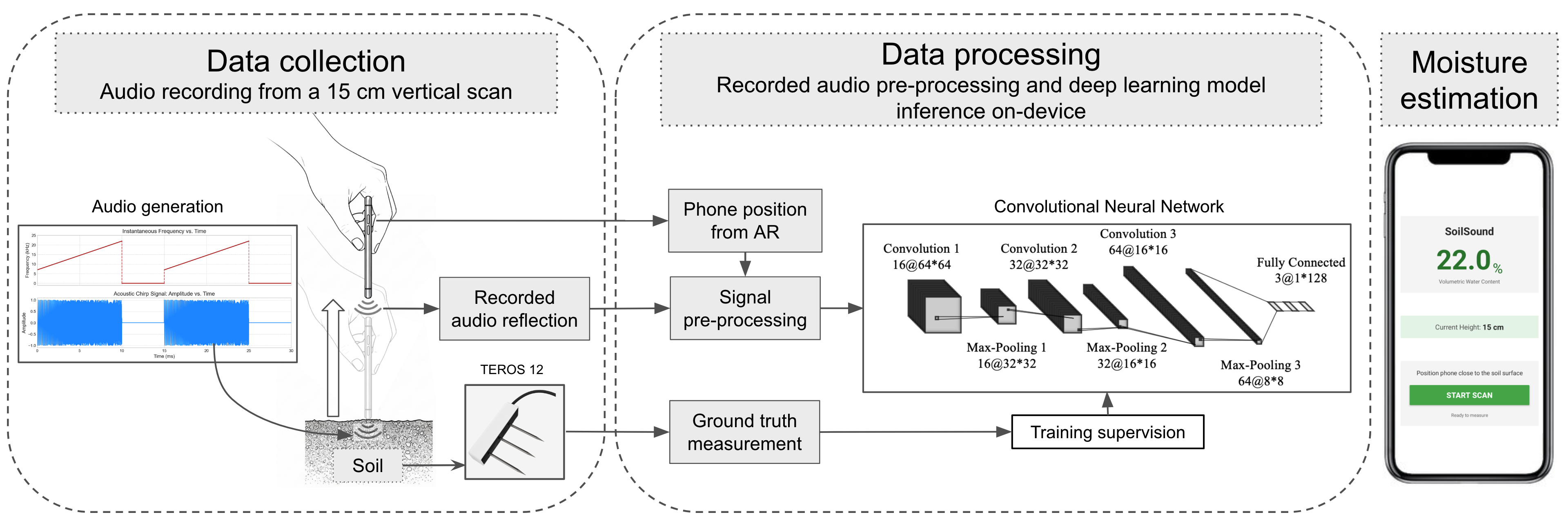}
    \caption{system overview}
    \vspace{-10pt} 
    \label{fig:system_overview}
\end{figure*}
\subsection{Overview}

\name{} works by sending FMCW acoustic chirp signal using a smartphone speaker into the soil, and recording the reflections from the soil surface using a (colocated) microphone in a vertical scan. Because of the vertical scan, the recording gives us FMCW acoustic reflections containing soil moisture information at different heights from the soil surface. After signal pre-processing, the recording is converted to a 2D range profile. A convolutional neural network is then used to estimate moisture from a 2D range profile. We explain each component of \name{} in detail below, and the system overview is shown in Figure \ref{fig:system_overview}.

\subsection{Data Acquisition}

\subsubsection{Acoustic Signal Generation}
The system continuously transmits 10~ms chirps with frequency ranging from 7 to 22 kHz from the smartphone speaker. We chose this frequency range as the environmental noise floor is lower in this range. In between the chirps, we insert a 5 ms silent gap which is used for synchronization in later steps.

\subsubsection{Position Tracking via ARCore}
When the user performs a vertical scan, the system needs to track the height of the smartphone from the soil surface. We use ARCore's visual-inertial odometry~\cite{GoogleARCore} to obtain the height. The system fuses data from the phone's IMU with visual features extracted from the camera feed to estimate 6-DOF pose at frame rate of 30 Hz. We extract the vertical component of position relative to an initial reference point established when scanning begins. This enables us to track the phone’s height throughout the scan while recording the corresponding acoustic reflections at each height. Approximately 100-200 chirps are collected during a typical 5-second scan, capturing the soil’s acoustic reflection across
multiple observation heights.

The position data serves two critical functions in our system. First, it provides real-time scanning quality feedback to ensure consistent data collection. The system monitors scanning velocity and suggests re-scanning if the speed falls outside the acceptable range of 1-5 cm/s (optimal: 3 cm/s) or exhibits excessive variation (velocity standard deviation > 2 cm/s). This quality control prevents corrupted scans resulting from inconsistent motion or hand tremor. Second, the position data determines when to terminate data collection—once the phone reaches 15 cm above the starting position. We stop at 15 cm height because the power of reflected signal becomes low beyond that point.

\subsection{Signal Processing Pipeline}
Once the vertical scan ends, the system stops transmitting chirps and begins pre-processing the recorded signal.

\subsubsection{Synchronization and Boundary Detection}
The recorded signal has a distinctive pattern, a periodic silence followed by a sharp energy onset, that serves as an unambiguous synchronization marker. When a sudden energy increase is detected after a quiet period, it is marked as the beginning of a new chirp in the received signal. 
To find the energy increase, we compute short-time energy using a sliding window:
\begin{equation}
E[n] = \sum_{k=0}^{W-1} x^2[n-k]
\end{equation}
where $W = 20$ samples (0.42 ms at 48 kHz sampling rate).

To find the beginning of the chirp, we search for positions where energy transitions from low to high—indicating the end of a gap and start of a chirp. Specifically, we identify points where the ratio between subsequent high-energy (chirp) and preceding low-energy (gap) regions exceeds 8:1. Once found, we verify this is a true chirp boundary by confirming that similar transitions occur at 720-sample intervals (corresponding to our 15 ms cycle period).
After detecting three consecutive cycles with consistent spacing, we achieve synchronization lock. From this point forward, we know exactly where each chirp begins in the received signal, enabling coherent processing of the echoes. This synchronization typically completes within 45 ms (three cycles) of data acquisition. This method of chirp boundary detection gives us better result than traditional correlation-based methods.

\subsubsection{Range Extraction and Direct Path Cancellation}
Each received chirp is first dechirped by multiplying it with the transmitted chirp's complex conjugate, converting time delays into intermediate beat frequencies proportional to height of the device from the soil surface. We then use blackman windowing function to reduce spectral leakage. We then apply a 1920 point discrete fourier transform (DFT), to obtain a range profile where each frequency bin maps to distance as $d = cf/(2B/T)$, yielding a range resolution of 1.14~cm. In the resulting range profile, the direct acoustic path from speaker to microphone creates a strong interference that masks soil reflections. To address this, we employ the template-based cancellation technique adapted from~\cite{sun2022aim} as follows: during initialization, we record a template by averaging 50 range profiles $(<1s)$ captured with the phone held away from reflective surfaces. During this step, we subtract this template from each measured range profile to remove the static direct path while preserving time-varying soil reflections. 

\subsubsection{Constructing 2D matrix with range profile for different heights}
The recording consists of chirps reflected during the entire vertical scan when the smartphone was moved to different heights. So we take the sequential range profiles obtained in the above step and directly assemble it into a 2D range profile image where each row represents a range profile at a specific height of the smartphone. 

For consistent neural network input, we resample this variable-sized image to a fixed 64×64 resolution. Depending upon the speed of the vertical scan, the recording could have 100-200 chirps at different heights. To ensure consistency, we uniformly sample 64 range profiles from the recording. Then each of that range profile is cropped to the first 64 bins as they contain the relevant information. This standardized 2D representation captures the complete acoustic response across multiple observation heights in a format suitable for deep learning.

\subsubsection{Mode-Based Noise Removal}

\begin{figure}[hbtp]
    \centering
    \includegraphics[width=0.9\linewidth]{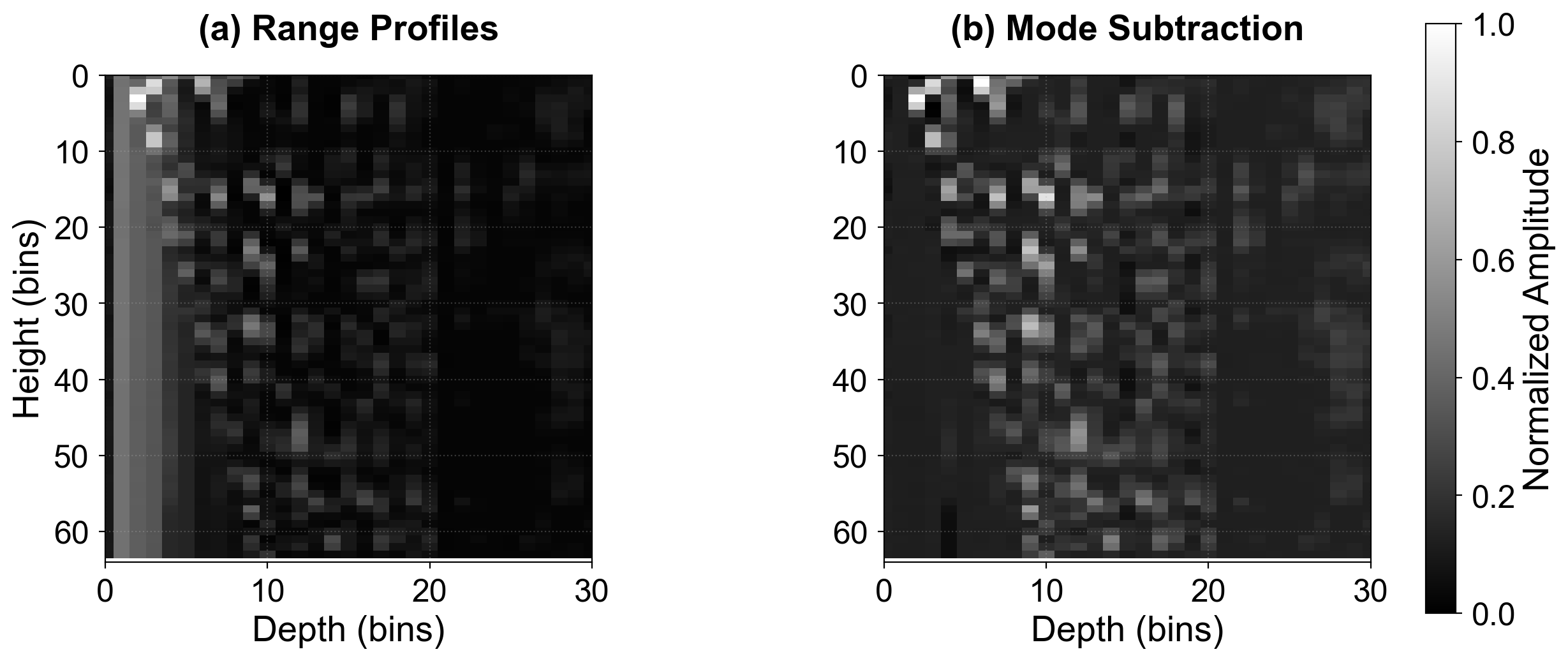}
    \caption{Effect of mode removal on the 2D range profile: (a) before subtraction, (b) after subtraction.}
    \vspace{-10pt} 
    \label{fig:preprocessing}
\end{figure}


In the 2D input, the range profiles at different heights still contain substantial background noise across different range bins, as shown in \cite{sun2022aim}. 
A key observation in our vertical scanning setup is that genuine soil reflections appear sparsely i.e for each height of the smartphone from the soil surface only few bins in the range profile contain actual reflections (either specular or scattering), while the majority capture background noise. To address this we perform a mode-base noise removal in the 2D range profile array. We compute the mode for each column i.e the most frequently occurring value, which represents the consistent noise floor at that range. Subtracting this mode from each column removes the static noise background while preserving genuine reflections that deviate from this baseline. After template removal, the 2D height-range image $S$ is cleaned as follows:
\begin{equation}
S_{\text{clean}}[:, j] = S[:, j] - \text{mode}(S[:, j])
\end{equation}
and then standardized using mean and standard deviation for the convolutional neural network.

As shown in Figure~\ref{fig:preprocessing}(b), this operation dramatically improves signal clarity, revealing soil reflections that were previously obscured by noise.

\subsection{Deep Learning Model for Moisture Estimation}
\subsubsection{Architecture}
The preprocessed 64×64 images are fed to a convolutional neural network that performs regression to estimate moisture. The CNN follows a standard architecture with three convolutional blocks, each consisting of convolution, batch normalization, and LeakyReLU activation ($\alpha=0.1$). The blocks progressively increase channel depth (1→16→32→64) while reducing spatial resolution through $2\times2$ max pooling. The feature maps are then flattened and passed through fully connected layers (4096→128→64→1) with dropout regularization, outputting a single moisture percentage.

\subsubsection{Optimization}
We train the model using multiple soil samples in leave-one-out cross-validation fashion. The error for each sample is a range-aware VWC error, as defined in CoMEt~\cite{khan2022estimating}, which accounts for ground truth measurement uncertainty. Contact-based moisture sensors provide point measurements that vary across measurements even at the same location. To account for this, we collect 5 ground truth measurements for each soil sample to establish a ground truth range $[\theta_{min}, \theta_{max}]$, and compute the VWC-error as:
\begin{equation}
\text{VWC-error} = 
\begin{cases}
0 & \text{if } \theta_{min} \leq \hat{\theta} \leq \theta_{max} \\
|\hat{\theta} - \theta_{max}| & \text{if } \hat{\theta} > \theta_{max} \\
|\theta_{min} - \hat{\theta}| & \text{if } \hat{\theta} < \theta_{min}
\end{cases}
\label{eq:range_mae}
\end{equation}
where $\hat\theta$ is the VWC soil moisture predicted by the model.

Finally, the model is then optimized using the mean squared VWC-error loss. This custom loss formulation recognizes that any prediction within the measurement uncertainty range is equally valid, preventing the model from overfitting to noise in individual sensor readings. 


\section{Smartphone app}

We implemented \name{} as an Android application running on a Samsung Galaxy S10e smartphone. 
The S10e smartphone has the necessary hardware capabilities: i)a speaker and microphone colocated at the bottom suitable for ground-directed acoustic sensing. It also gives access to automatic gain control (AGC) of the microphone. ii) an xx camera for ARCore visual-inertial tracking. The phone's Snapdragon 855 processor enables real-time signal processing and neural network inference on-device. The app's user interface (UI) and implementation details are provided below.

\textbf{User Interface:}
The app consists an interface that guides the users through the scanning process with real-time visual feedback. During scanning, a height indicator shows the current position of the smartphone relative to the target 15 cm range, with color coding (green for proper speed, yellow for too slow, red for too fast) to help maintain a 3 cm/s velocity. 
After scanning, the app displays the estimated moisture percentage (volumetric water content).
Users can save results with GPS coordinates and timestamps for field mapping applications. A sample UI is shown in Figure \ref{fig:system_overview}.


\textbf{Software:} The smartphone app was written in Kotlin programming language and built using the android studio. We use the AudioTrack and AudioRecord APIs for acoustic signal generation and capture, and ARCore API for spatial tracking.

\textbf{Training: }In the training phase, raw data is transmitted to a connected laptop (MacBook Pro M1) for processing. The app streams data to a laptop through TCP connection over ADB reverse tunneling. The deep learning model is trained using PyTorch on an NVIDIA RTX 3090 GPU. The trained models are exported to TorchScript format and quantized for mobile deployment. This quantized model is then stored in the phone and can easily run on the S10e's CPU.

\textbf{Inference:} For inference, we developed a standalone Android application that loads the trained model directly on the device using PyTorch Mobile runtime. We implement the preprocessing pipeline in Kotlin. The app performs real-time FMCW processing and uses the stored model  to predict the soil moisture level entirely on-device.

\section{Evaluation}
We evaluate \name{}’s overall performance for soil moisture estimation by conducting multiple experiments across different soil types and different environments. We perform a small user study to test the usability of our system. Finally we conduct an outdoor field study to show \name{}'s feasibility to measure moisture of the soil in the real world using a smartphone app without disturbing the soil. In this section, we describe in detail, the various experiment setup, the evaluation metrics and the results. We begin by defining soil moisture, describing the ground truth measurement device, and lastly our evaluation metrics.

\textbf{Soil moisture definition:}
In this work we use the volumetric water content (VWC) to define the soil moisture, which is computed as the ratio of the volume of water to the total volume of soil computed as following and reported as a percentage:
\begin{equation}
\theta_{v}={V_{\text{water}}}/{V_{\text{soil}}}
\end{equation}

\textbf{Ground truth measurement device:}
We estimated the ground truth VWC of the soil using the AROYA Solus 3-in-1 Soil Analyzer \cite{noauthor_aroya_nodate}, which has a capacitance-based TEROS 12 soil moisture sensor \cite{metergroup_teros12}.
It has an epoxy-filled head with three 5.5 cm long stainless steel rods arranged in a linear array that are inserted into the soil to measure moisture at a
desired depth. The TEROS 12 is a capacitance-based sensor, detecting changes in the apparent bulk dielectric permittivity, which is strongly influenced by
water content. Across both laboratory and field validations in various mineral soils, a linear calibration model for the sensor resulted in an MAE of 3\% \cite{cominelli2024calibration}.
As a contact-based sensor, the TEROS 12 provides point measurements that reflect soil moisture content at the specific location where it is installed. As the ground truth moisture measurements vary across locations, instead of measuring soil moisture at a single location, measurements were taken at five nearby locations near the surface to obtain a representative ground truth soil moisture ground truth range.

\textbf{Evaluation Metric.}
 To evaluate the performance of \name{}, we compare the soil moisture estimated by our model with the soil moisture range provided by the ground truth sensor. Specifically, we compute the range-aware VWC error as defined in Equation \ref{eq:range_mae}. To ensure robust measurements, we perform multiple scans for each soil sample in all the experiments. The VWC-error is individually calculated for each phone scan of each soil sample. We then report the mean absolute error (MAE) of the range-aware VWC error for each soil sample.

\subsection{In-lab testing}
to test the accuracy of \name{}, we first performed a controlled experiment in a lab setting with boxes of specially curated soil .
\textbf{Soil Preparation:} Our laboratory soil preparation aimed to generate samples representing real-world soil conditions with diverse moisture levels. We prepared two base soil types: (1) loamy sand collected locally from our city, classified as low in organic matter, and (2) commercial potting soil purchased from Amazon, representing typical home and garden use with high organic matter content.

We began by drying 200L of each soil type in an industrial oven. We then split each of the soil type into nine 1~kg samples. To each of these split sample, we systematically added water in different increments of 62.5mL (62.5mL, 125mL, 187.5mL, etc.). This leads to nine soil samples with different moisture levels. After adding water we mixed each sample thoroughly by hand (wearing gloves) for 5 minutes to ensure homogeneity. We then transferred the sample to 1.9L IKEA Pruta containers filling it to capacity. Then samples were set for 48 hours before testing. Prior to each experiment, we verified the moisture content of the sample using a Taro probe. The final moisture ranges achieved were 12.71\% to 35.94\% for the loamy sand and 8.48\% to 32.85\% for the high-organic potting soil.

\textbf{Acoustic data collection:} After preparing the soil samples, we take approximately 15 vertical scans per soil sample, which are individual data points for our experiment. Immediately after scanning, we use our ground truth measurement device to capture the most accurate ground truth range. 

\subsubsection{Estimating soil moisture for single soil type} First, we evaluate the accuracy of soil moisture measurement when system is trained and tested on the same soil type. For this experiment, we use the loamy soil samples and measure the accuracy using the LOOCV method. In each fold we train on $n-1$ soil samples' scans, and test on the remaining one sample's scans to compute the range-aware MAE for the latter soil sample. This process is repeated for all nine soil samples.

\begin{figure}[hbtp]
    \vspace{-10pt} 
    \centering
    \includegraphics[width=\linewidth]{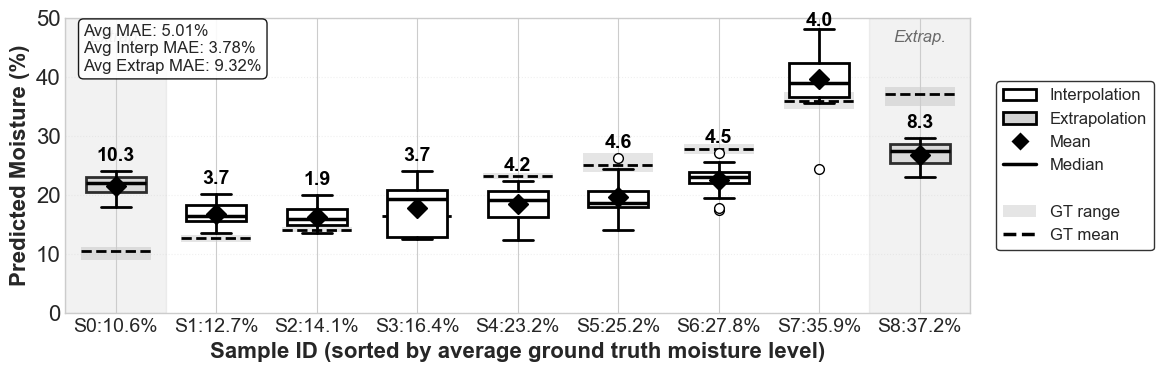}
    \caption{LOOCV soil moisture estimation errors for 9 lab-prepared soil samples.}
        \vspace{-10pt} 
    \label{fig:lomlo_results}
\end{figure}

The results of the LOOCV is reported in Figure \ref{fig:lomlo_results}, across 9 soil samples corresponding to moisture levels from 12.7\% to 37.2\% VWC.

Our model achieves an overall MAE of \labMAE{} , with 3.78\% on interpolation tasks (moisture levels within the training range) and 9.32\% on extrapolation tasks (moisture levels outside the training range). This performance gap is expected, as generalizing beyond the training distribution is inherently challenging for deep learning models. In practical deployment, training on the full moisture range eliminates this extrapolation error. 


\textbf{Training model:} At the end of above experiment, we take all the loamy sand soil samples with different moisture levels to train a model. We then freeze this model and use it for training for all the subsequent experiments in this section.

\subsubsection{Performance on Unseen Soil Types}

\begin{figure}[ht]
    \vspace{-10pt} 
    \centering
    \includegraphics[width=\linewidth]{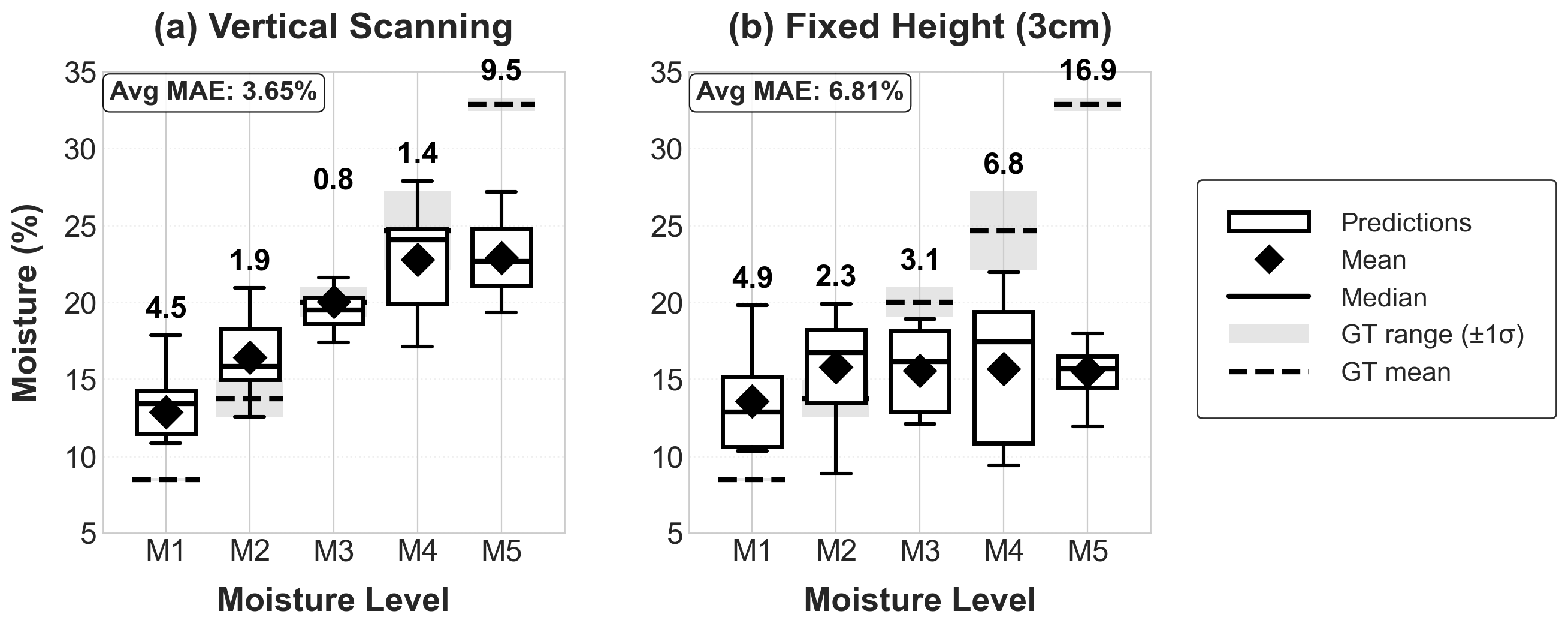}
    \caption{Model performance on organic potting soil (unseen soil type) after training exclusively on loamy sand.}
    \vspace{-10pt} 
    \label{fig:eval_unseen_ablation}
\end{figure}

We then evaluate the generalizability of our model for different soil types. In this experiment we train the model using the loamy soil samples and test on organic potting soil samples with high organic matter content. It has a fundamentally different soil composition from the previous experiment's loamy sand soil, and is commonly used in urban gardening. Figure~\ref{fig:eval_unseen_ablation}(a) presents the results across five unseen potting soil samples with five different moisture levels.
The model demonstrates strong generalization capability, achieving high accuracy at intermediate moisture levels with acceptable performance even at extremes (4.5\% error at dry conditions, 10.0\% at near-saturation). However we also note that the model struggles in the extreme cases similar to the in-lab experiments.

This experiment shows our model's ability to generalize across fundamentally different soil types without any fine-tuning or calibration; Our model effectively captures moisture-invariant features. The slightly degraded performance at extreme moisture conditions (very dry and saturated) is expected as extrapolation might happen at a different range for different soil types. 

\subsection{Benchmarks: } The above experiments show that \name{} can estimate moisture of soil sample by performing vertical scans of smartphone and estimating the range profile of the reflected signals at these different heights. To further verify this we perform two benchmark experiments. 
\subsubsection{Vertical Scanning Ablation}
First we conduct an ablation study to evaluate the impact of the vertical scan mechanism in \name{}. Instead of using the model trained with multiple scans on loamy soil samples, we train a separate fixed height model for the same soil samples. In this model, instead of a full vertical scan, we only use data from a single height of the smartphone from the soil namely 3~cm. While both models' losses converged during training, their generalization capabilities differed dramatically.

When we tested this model on the unseen organic soil samples from previous experiment, the fixed-height model failed catastrophically: it produced nearly constant predictions around 16\% VWC regardless of actual moisture content (8.5\%--32.9\%) (Figure~\ref{fig:eval_unseen_ablation}(b)).

This failure reveals that single-height measurements capture soil-specific acoustic signatures rather than moisture-invariant features and overfits the model. Vertical scanning provides essential geometric diversity, with samples from different height enabling robust moisture estimation across soil types.

\subsubsection{Model Interpretability:} 
To interpret the model trained by our system, we visualize the activation map from the output of the first convolutional layer when processing an unseen soil sample. 

\begin{figure}[h]
    \vspace{-10pt} 
    \centering
    \includegraphics[width=0.2\textwidth]{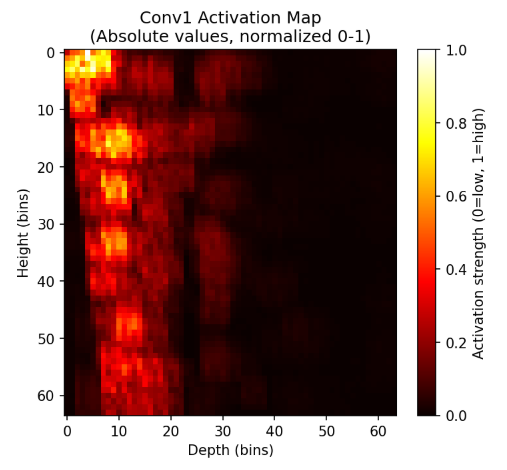}
    \caption{First convolutional layer activation map showing learned features from a soil moisture scan.}
    \vspace{-10pt} 
    \label{fig:activation_map}
\end{figure}

Figure~\ref{fig:activation_map} displays the activation map and it shows that the model discovers the direct surface reflection path (left edge, corresponding to the shortest acoustic path), with particularly high activation strength at lower heights where the phone is closer to the soil surface. Additionally, the model exhibits meaningful activations at extended range bins (deeper into the image), indicating it learns to utilize diffuse scattering information from oblique angles. These findings are consistent with our soil acoustic reflection model from Section 3.

\subsection{Usability Study}
As shown above, the vertical scanning of the smartphone is essential for the working of \name{}. However, this process could be tricky as there might be significant variance in scanning speed and direction between subjects. To test that the system can work in a robust manner for different subjects, we conduct a small user study with five subjects to test the usability of the system. These users have no prior experience with \name{} and are only shown a brief demonstration of vertical scanning at 3 cm/s at the beginning of the study. During the study they performed multiple vertical scans to estimate the moisture of the organic soil with the Ground truth moisture of 19.0\% $\pm$ 2.29\%. 

\begin{figure}[h]
    \vspace{-10pt} 
    \centering
    \includegraphics[width=0.9\linewidth]{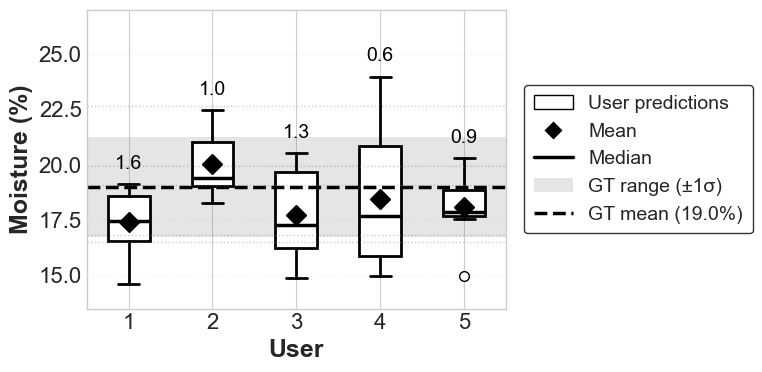}
    \caption{Variation of moisture estimation across five users on a single unseen soil sample.}
    \vspace{-10pt} 
    \label{fig:usability}
\end{figure}

Figure~\ref{fig:usability} plots the estimated moisture of the sample for all the five users. The fig shows that \name{} sustained an average range-aware MAE of 1.14\% across the five users with max MAE of 1.6\% for first user. This confirms that \name{} is usable in the real world and is robust to minor variations in the scanning process across users. Further users also noted that the non-invasive measurement is a key advantage over traditional sensors that require soil penetration, which can cause damage to plant roots. These results validate practical deployment potential for all users.

\begin{figure*}[]
    \vspace{-10pt} 
    \centering
    \includegraphics[width=\textwidth]{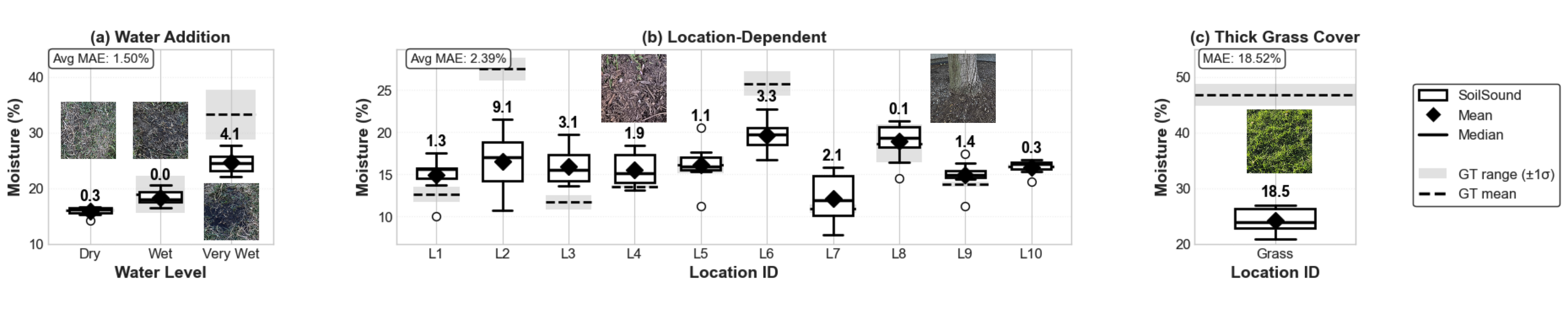}
    \vspace{-20pt} 
    \caption{Comprehensive outdoor field evaluation of \name{}. (a) Moisture responsiveness test. (b) Location-dependent evaluation: Model performance across 10 outdoor locations with varying soil conditions. (c) Sample location with thick grass coverage.}
    \vspace{-10pt} 
    \label{fig:eval_outdoor}
\end{figure*}

\subsection{Outdoor Field Evaluation}

Finally, we conducted three different outdoor experiments to validate \name{} s ability to measure soil moisture in outdoor fields in the real world. 

\subsubsection{outdoor moisture responsiveness test}
First, we choose one outdoor location and we progressively increased the moisture content by adding water between measurements. In each instance, we measured with both groundtruth and smartphone using SoilSound. Figure~\ref{fig:eval_outdoor} shows three measurement sessions at the same location: initially dry conditions (15.9\% VWC), after moderate watering (18.9\% VWC), and after substantial watering resulting in near-saturation (33.3\% VWC). The visible darkening of the soil in the photographs confirms the increasing moisture levels.

The system demonstrates good accuracy across the full moisture range at this single spot. At lower to moderate moisture levels (15.9\% and 18.9\% VWC), the model achieves near-perfect performance (0.3\% and 0.0\% error, respectively) and natural soil heterogeneity. At high moisture content (33.33\% VWC), the error increases to 4.1\%, which shows the same effect of performance decaying at high moisture shown in in-lab study

These results validate that our system can accurately track moisture changes in the outdoor field soil over time, a critical requirement for irrigation management and environmental monitoring applications.





\subsubsection{location-dependent test}

To assess the system's performance across diverse field conditions, we evaluated \name{} at 10 distinct outdoor locations (L1 to L10) within our city, with naturally varying soil compositions, moisture contents, and surface conditions. Specifically the locations L1, L2, L8, and L9 are closer a plant. We chose these locations to test the feasibility of using \name{} to determine whether a plant needs watering. At each location we performed multiple ground truth measurements using the TEROS-12 sensor and vertical scans using the smartphone. 

As shown in Figure~\ref{fig:eval_outdoor}(b), the model achieved an average MAE of 2.39\% across  10 locations. Nine locations (L1, L3-L10) with ground truth moisture ranging from 10.9\% to 25.8\% achieved excellent performance with errors between 0.1\% and 3.3\%. Only Location L2 with high moisture content (27.47\% VWC) increased our system error to 9.1\%. The consistent performance across diverse locations with a variety of soil conditions validates the model's ability to generalize across the wide range of conditions encountered in real-world agricultural and environmental monitoring applications.

\subsubsection{Grass coverage test}
To evaluate system performance under grass coverage, we conducted moisture estimation experiments on areas with thick grass cover. The results shown in Figure~\ref{fig:eval_outdoor}(c) indicate that while the actual moisture level beneath the grass measured 46.73\% VWC, our system consistently underestimated these values, yielding an MAE of 18.52\%. This demonstrates a key limitation of our system when estimating moisture through heavy grass coverage. The underestimation occurs because grass itself contains significant moisture, and its structured surface geometry differs substantially from bare soil, causing our acoustic model to produce inaccurate predictions in these conditions. One potential solution is to build a separate model for acoustic reflections of surafces covered by grass.

\subsection{On-device Performance}

Finally, We evaluated \name{} s performance on the Samsung Galaxy S10e smartphone and Table~\ref{tab:mobile_performance} presents the performance metrics averaged over 5 scans.




\begin{table}[h]
\centering
\caption{Mobile Inference Performance (Samsung Galaxy S10e)}
\small
\begin{tabular}{ccc|c|c}
\hline
\textbf{Pre-processing} & \textbf{Inference} & \textbf{Total} & \textbf{Model} & \textbf{Energy} \\
\hline
$720{\pm}95$ ms & $25{\pm}15$ ms & $745{\pm}89$ ms & 2.2 MB & $<$0.01 mJ \\
\hline
\end{tabular}
\label{tab:mobile_performance}
\vspace{-10pt} 
\end{table}

The complete measurement process, from completing a 5-second vertical scan (15 cm height at 3 cm/s average velocity) to displaying the moisture result, takes under 1 second of processing time. The preprocessing stage (720 ms) dominates the computational cost, including synchronization, dechirping, FFT, and 2D image construction. The CNN inference itself requires only 25 ms, demonstrating the efficiency of our quantized model. With a model size of 2.2 MB and negligible energy consumption per measurement, SoilSound imposes minimal resource requirements on COTS smartphones. This efficiency enables real-time moisture feedback during field operations without draining battery or storage resources.

\section{Limitations and Discussion}

\name{} shows a pathway to enable smartphone based soil moisture sensing in the real world. In this section, we discuss some of the current limitations and potential future directions of our system.

\textbf{Depth Penetration Limitations.} 
The primary limitation of \name{} lies in its limited penetration depth. The moisture content value obtained with Soilsound represents that of the surface soil (~a few cm) in a small area (<1 $m^2$). As discussed in Section~3, transmission coefficient from air to soil is very low, therefore deeper soil layers remain effectively "invisible" to \name{} due to the fundamental physics of sound propagation. 

\textbf{Variable soil compositions and conditions.} 
While our evaluation already included diverse soil types across approximately 150 sample scans, the natural variability of soil compositions and conditions are much larger and can be a challenge for \name{}. For example, soils with strong grass cover, high salt contents, high organic matter concentrations, frozen, or with standing water may violate the assumptions for acoustic reflections underlying our model. The specular/scattering ratio that enables moisture detection could be dominated by other factors in these cases. In the future, this can be potentially addressed by enabling large scale studies that includes diverse conditions.

\textbf{Towards Community-Driven Refinement.} 
Rather than attempting exhaustive data collection across all possible soil compositions and conditions by ourselves, we envision the system improving through distributed, community-driven refinement. Deploying the system by a range of collaborators (e.g., researchers, practitioners) as a practical application enables crowdsourced data collection, where users optionally contribute anonymized scans with contextual metadata (geographic location, soil classification, precipitation history) in addition to moisture data collected with their own tools (i.e., with a different method). This approach transforms the scaling challenge into an opportunity to improve the model for subsequent users. However, this approach will have some engineering challenges such as cross-platform deployment (currently limited to Android devices), privacy-preserving data aggregation, and federated learning architectures that balance model improvement with user data protection. The transition from research prototype to community-scale deployment will require addressing these sociotechnical considerations alongside the core sensing capabilities.

\section{Conclusion}

\name{} represents the first smartphone-based acoustic soil moisture sensing system that requires no additional hardware and no disturbance of soil. We modeled soil's moisture level based on acoustic reflective properties and designed a vertical scanning method that collect independent information of the soil at different heights above the soil.  Our CNN-based approach learns to extract moisture-relevant information from the 2D range profile representations, allowing soil moisture measurement without explicit field calibration, achieving \outdoorMAE{} MAE on 10 outdoor field samples.




\bibliographystyle{ACM-Reference-Format}
\bibliography{ref}


\begin{thebibliography}{36}


\ifx \showCODEN    \undefined \def \showCODEN     #1{\unskip}     \fi
\ifx \showDOI      \undefined \def \showDOI       #1{#1}\fi
\ifx \showISBNx    \undefined \def \showISBNx     #1{\unskip}     \fi
\ifx \showISBNxiii \undefined \def \showISBNxiii  #1{\unskip}     \fi
\ifx \showISSN     \undefined \def \showISSN      #1{\unskip}     \fi
\ifx \showLCCN     \undefined \def \showLCCN      #1{\unskip}     \fi
\ifx \shownote     \undefined \def \shownote      #1{#1}          \fi
\ifx \showarticletitle \undefined \def \showarticletitle #1{#1}   \fi
\ifx \showURL      \undefined \def \showURL       {\relax}        \fi
\providecommand\bibfield[2]{#2}
\providecommand\bibinfo[2]{#2}
\providecommand\natexlab[1]{#1}
\providecommand\showeprint[2][]{arXiv:#2}

\bibitem[noa({[n.\,d.]})]%
        {noauthor_aroya_nodate}
 \bibinfo{year}{[n.\,d.]}\natexlab{}.
\newblock \bibinfo{title}{{Aroya} {Solus} 3 {in} 1 {Bluetooth} {EC} {Sensor}, {Soil} {Moisture} {Soil} {Temperature} - {Water} {Content} {Sensor}}.
\newblock
\newblock
\urldef\tempurl%
\url{https://forevergreenindoors.com/products/solus-bluetooth-water-content-sensor}
\showURL{%
\tempurl}


\bibitem[met({[n.\,d.]})]%
        {metergroup_teros12}
 \bibinfo{year}{[n.\,d.]}\natexlab{}.
\newblock \bibinfo{title}{{TEROS} 12 Advanced Soil Moisture Sensing}.
\newblock
\newblock
\urldef\tempurl%
\url{https://metergroup.com/products/teros-12/}
\showURL{%
\tempurl}


\bibitem[Adamo et~al\mbox{.}(2004)]%
        {adamo2004acoustic}
\bibfield{author}{\bibinfo{person}{Francesco Adamo}, \bibinfo{person}{Gregorio Andria}, \bibinfo{person}{Filippo Attivissimo}, {and} \bibinfo{person}{Nicola Giaquinto}.} \bibinfo{year}{2004}\natexlab{}.
\newblock \showarticletitle{An acoustic method for soil moisture measurement}.
\newblock \bibinfo{journal}{\emph{IEEE transactions on instrumentation and measurement}} \bibinfo{volume}{53}, \bibinfo{number}{4} (\bibinfo{year}{2004}), \bibinfo{pages}{891--898}.
\newblock


\bibitem[Allmaras et~al\mbox{.}(1966)]%
        {allmaras1966total}
\bibfield{author}{\bibinfo{person}{RR Allmaras}, \bibinfo{person}{Robert~E Burwell}, \bibinfo{person}{William~E Larson}, \bibinfo{person}{Robert~F Holt}, {and} \bibinfo{person}{WW Nelson}.} \bibinfo{year}{1966}\natexlab{}.
\newblock \showarticletitle{Total porosity and random roughness of the interrow zone as influenced by tillage}.
\newblock \bibinfo{journal}{\emph{Conservation research report}} (\bibinfo{year}{1966}).
\newblock


\bibitem[Beckmann and Spizzichino(1987)]%
        {beckmann1987scattering}
\bibfield{author}{\bibinfo{person}{Petr Beckmann} {and} \bibinfo{person}{Andre Spizzichino}.} \bibinfo{year}{1987}\natexlab{}.
\newblock \showarticletitle{The scattering of electromagnetic waves from rough surfaces}.
\newblock \bibinfo{journal}{\emph{Norwood}} (\bibinfo{year}{1987}).
\newblock


\bibitem[Chang et~al\mbox{.}(2022)]%
        {chang2022sensor}
\bibfield{author}{\bibinfo{person}{Zhaoxin Chang}, \bibinfo{person}{Fusang Zhang}, \bibinfo{person}{Jie Xiong}, \bibinfo{person}{Junqi Ma}, \bibinfo{person}{Beihong Jin}, {and} \bibinfo{person}{Daqing Zhang}.} \bibinfo{year}{2022}\natexlab{}.
\newblock \showarticletitle{Sensor-free soil moisture sensing using lora signals}.
\newblock \bibinfo{journal}{\emph{Proceedings of the ACM on Interactive, Mobile, Wearable and Ubiquitous Technologies}} \bibinfo{volume}{6}, \bibinfo{number}{2} (\bibinfo{year}{2022}), \bibinfo{pages}{1--27}.
\newblock


\bibitem[Cihlar and Ulaby(1974)]%
        {cihlar1974dielectric}
\bibfield{author}{\bibinfo{person}{Josef Cihlar} {and} \bibinfo{person}{Fawwaz~Tayssir Ulaby}.} \bibinfo{year}{1974}\natexlab{}.
\newblock \bibinfo{booktitle}{\emph{Dielectric properties of soils as a function of moisture content}}.
\newblock \bibinfo{type}{{T}echnical {R}eport}.
\newblock


\bibitem[Cominelli et~al\mbox{.}(2024)]%
        {cominelli2024calibration}
\bibfield{author}{\bibinfo{person}{Sofia Cominelli}, \bibinfo{person}{Leonardo~D Rivera}, \bibinfo{person}{William~G Brown}, \bibinfo{person}{Tyson~E Ochsner}, {and} \bibinfo{person}{Andres Patrignani}.} \bibinfo{year}{2024}\natexlab{}.
\newblock \showarticletitle{Calibration of TEROS 10 and TEROS 12 electromagnetic soil moisture sensors}.
\newblock \bibinfo{journal}{\emph{Soil Science Society of America Journal}} \bibinfo{volume}{88}, \bibinfo{number}{6} (\bibinfo{year}{2024}), \bibinfo{pages}{2104--2122}.
\newblock


\bibitem[Darmon et~al\mbox{.}(2020)]%
        {darmon2020acoustic}
\bibfield{author}{\bibinfo{person}{Michel Darmon}, \bibinfo{person}{Vincent Dorval}, {and} \bibinfo{person}{Fran{\c{c}}ois Baqu{\'e}}.} \bibinfo{year}{2020}\natexlab{}.
\newblock \showarticletitle{Acoustic scattering models from rough surfaces: a brief review and recent advances}.
\newblock \bibinfo{journal}{\emph{Applied Sciences}} \bibinfo{volume}{10}, \bibinfo{number}{22} (\bibinfo{year}{2020}), \bibinfo{pages}{8305}.
\newblock


\bibitem[Ding and Chandra(2019)]%
        {ding2019towards}
\bibfield{author}{\bibinfo{person}{Jian Ding} {and} \bibinfo{person}{Ranveer Chandra}.} \bibinfo{year}{2019}\natexlab{}.
\newblock \showarticletitle{Towards low cost soil sensing using Wi-Fi}. In \bibinfo{booktitle}{\emph{The 25th annual international conference on mobile computing and networking}}. \bibinfo{pages}{1--16}.
\newblock


\bibitem[Ding et~al\mbox{.}(2023)]%
        {ding2023soil}
\bibfield{author}{\bibinfo{person}{Rong Ding}, \bibinfo{person}{Haiming Jin}, \bibinfo{person}{Dong Xiang}, \bibinfo{person}{Xiaocheng Wang}, \bibinfo{person}{Yongkui Zhang}, \bibinfo{person}{Dingman Shen}, \bibinfo{person}{Lu Su}, \bibinfo{person}{Wentian Hao}, \bibinfo{person}{Mingyuan Tao}, \bibinfo{person}{Xinbing Wang}, {et~al\mbox{.}}} \bibinfo{year}{2023}\natexlab{}.
\newblock \showarticletitle{Soil moisture sensing with uav-mounted ir-uwb radar and deep learning}.
\newblock \bibinfo{journal}{\emph{Proceedings of the ACM on Interactive, Mobile, Wearable and Ubiquitous Technologies}} \bibinfo{volume}{7}, \bibinfo{number}{1} (\bibinfo{year}{2023}), \bibinfo{pages}{1--25}.
\newblock


\bibitem[Entekhabi et~al\mbox{.}(2010)]%
        {entekhabi2010soil}
\bibfield{author}{\bibinfo{person}{Dara Entekhabi}, \bibinfo{person}{Eni~G Njoku}, \bibinfo{person}{Peggy~E O'neill}, \bibinfo{person}{Kent~H Kellogg}, \bibinfo{person}{Wade~T Crow}, \bibinfo{person}{Wendy~N Edelstein}, \bibinfo{person}{Jared~K Entin}, \bibinfo{person}{Shawn~D Goodman}, \bibinfo{person}{Thomas~J Jackson}, \bibinfo{person}{Joel Johnson}, {et~al\mbox{.}}} \bibinfo{year}{2010}\natexlab{}.
\newblock \showarticletitle{The soil moisture active passive (SMAP) mission}.
\newblock \bibinfo{journal}{\emph{Proc. IEEE}} \bibinfo{volume}{98}, \bibinfo{number}{5} (\bibinfo{year}{2010}), \bibinfo{pages}{704--716}.
\newblock


\bibitem[Feng et~al\mbox{.}(2022)]%
        {feng2022lte}
\bibfield{author}{\bibinfo{person}{Yuda Feng}, \bibinfo{person}{Yaxiong Xie}, \bibinfo{person}{Deepak Ganesan}, {and} \bibinfo{person}{Jie Xiong}.} \bibinfo{year}{2022}\natexlab{}.
\newblock \showarticletitle{Lte-based low-cost and low-power soil moisture sensing}. In \bibinfo{booktitle}{\emph{Proceedings of the 20th ACM Conference on Embedded Networked Sensor Systems}}. \bibinfo{pages}{421--434}.
\newblock


\bibitem[{Google}(2024)]%
        {GoogleARCore}
\bibfield{author}{\bibinfo{person}{{Google}}.} \bibinfo{year}{2024}\natexlab{}.
\newblock \bibinfo{title}{{ARCore: Google's Platform for Building Augmented Reality Experiences}}.
\newblock \bibinfo{howpublished}{\url{https://developers.google.com/ar}}.
\newblock
\newblock
\shownote{Accessed: 2025-09-01}.


\bibitem[Jiao et~al\mbox{.}(2023)]%
        {jiao2023soiltag}
\bibfield{author}{\bibinfo{person}{Wenli Jiao}, \bibinfo{person}{Ju Wang}, \bibinfo{person}{Yelu He}, \bibinfo{person}{Xiangdong Xi}, {and} \bibinfo{person}{Fuwei Wang}.} \bibinfo{year}{2023}\natexlab{}.
\newblock \showarticletitle{SoilTAG: Fine-grained soil moisture sensing through chipless tags}.
\newblock \bibinfo{journal}{\emph{IEEE Transactions on Mobile Computing}} \bibinfo{volume}{23}, \bibinfo{number}{3} (\bibinfo{year}{2023}), \bibinfo{pages}{2153--2170}.
\newblock


\bibitem[Jorapur et~al\mbox{.}(2015)]%
        {jorapur2015low}
\bibfield{author}{\bibinfo{person}{Nikhil Jorapur}, \bibinfo{person}{Vinay~S Palaparthy}, \bibinfo{person}{Shahbaz Sarik}, \bibinfo{person}{Jobish John}, \bibinfo{person}{Maryam~Shojaei Baghini}, {and} \bibinfo{person}{GK Ananthasuresh}.} \bibinfo{year}{2015}\natexlab{}.
\newblock \showarticletitle{A low-power, low-cost soil-moisture sensor using dual-probe heat-pulse technique}.
\newblock \bibinfo{journal}{\emph{Sensors and Actuators A: Physical}}  \bibinfo{volume}{233} (\bibinfo{year}{2015}), \bibinfo{pages}{108--117}.
\newblock


\bibitem[Josephson et~al\mbox{.}(2021)]%
        {josephson2021low}
\bibfield{author}{\bibinfo{person}{Colleen Josephson}, \bibinfo{person}{Manikanta Kotaru}, \bibinfo{person}{Keith Winstein}, \bibinfo{person}{Sachin Katti}, {and} \bibinfo{person}{Ranveer Chandra}.} \bibinfo{year}{2021}\natexlab{}.
\newblock \showarticletitle{Low-cost in-ground soil moisture sensing with radar backscatter tags}. In \bibinfo{booktitle}{\emph{Proceedings of the 4th ACM SIGCAS Conference on Computing and Sustainable Societies}}. \bibinfo{pages}{299--311}.
\newblock


\bibitem[Khan and Shahzad(2022)]%
        {khan2022estimating}
\bibfield{author}{\bibinfo{person}{Usman~Mahmood Khan} {and} \bibinfo{person}{Muhammad Shahzad}.} \bibinfo{year}{2022}\natexlab{}.
\newblock \showarticletitle{Estimating soil moisture using RF signals}. In \bibinfo{booktitle}{\emph{Proceedings of the 28th annual international conference on mobile computing and networking}}. \bibinfo{pages}{242--254}.
\newblock


\bibitem[Kiv et~al\mbox{.}(2022)]%
        {kiv2022smol}
\bibfield{author}{\bibinfo{person}{Daniel Kiv}, \bibinfo{person}{Garvita Allabadi}, \bibinfo{person}{Berkay Kaplan}, {and} \bibinfo{person}{Robin Kravets}.} \bibinfo{year}{2022}\natexlab{}.
\newblock \showarticletitle{Smol: Sensing soil moisture using LoRa}. In \bibinfo{booktitle}{\emph{Proceedings of the 1st ACM Workshop on No Power and Low Power Internet-of-Things}}. \bibinfo{pages}{21--27}.
\newblock


\bibitem[Kodikara et~al\mbox{.}(2014)]%
        {kodikara2014soil}
\bibfield{author}{\bibinfo{person}{Jayantha Kodikara}, \bibinfo{person}{Pathmanathan Rajeev}, \bibinfo{person}{Derek Chan}, {and} \bibinfo{person}{Chaminda Gallage}.} \bibinfo{year}{2014}\natexlab{}.
\newblock \showarticletitle{Soil moisture monitoring at the field scale using neutron probe}.
\newblock \bibinfo{journal}{\emph{Canadian Geotechnical Journal}} \bibinfo{volume}{51}, \bibinfo{number}{3} (\bibinfo{year}{2014}), \bibinfo{pages}{332--345}.
\newblock


\bibitem[Ledieu et~al\mbox{.}(1986)]%
        {ledieu1986method}
\bibfield{author}{\bibinfo{person}{J Ledieu}, \bibinfo{person}{P De~Ridder}, \bibinfo{person}{P De~Clerck}, {and} \bibinfo{person}{S Dautrebande}.} \bibinfo{year}{1986}\natexlab{}.
\newblock \showarticletitle{A method of measuring soil moisture by time-domain reflectometry}.
\newblock \bibinfo{journal}{\emph{Journal of Hydrology}} \bibinfo{volume}{88}, \bibinfo{number}{3-4} (\bibinfo{year}{1986}), \bibinfo{pages}{319--328}.
\newblock


\bibitem[Lu and Likos(2006)]%
        {lu2006suction}
\bibfield{author}{\bibinfo{person}{Ning Lu} {and} \bibinfo{person}{William~J Likos}.} \bibinfo{year}{2006}\natexlab{}.
\newblock \showarticletitle{Suction stress characteristic curve for unsaturated soil}.
\newblock \bibinfo{journal}{\emph{Journal of geotechnical and geoenvironmental engineering}} \bibinfo{volume}{132}, \bibinfo{number}{2} (\bibinfo{year}{2006}), \bibinfo{pages}{131--142}.
\newblock


\bibitem[Meisami-asl et~al\mbox{.}(2013)]%
        {meisami2013site}
\bibfield{author}{\bibinfo{person}{Elham Meisami-asl}, \bibinfo{person}{A Sharifi}, \bibinfo{person}{Hossein Mobli}, \bibinfo{person}{Afshin Eyvani}, {and} \bibinfo{person}{Reza Alimardani}.} \bibinfo{year}{2013}\natexlab{}.
\newblock \showarticletitle{On-site measurement of soil moisture content using an acoustic system}.
\newblock \bibinfo{journal}{\emph{Agricultural Engineering International: CIGR Journal}} \bibinfo{volume}{15}, \bibinfo{number}{4} (\bibinfo{year}{2013}), \bibinfo{pages}{1--8}.
\newblock


\bibitem[Oelze et~al\mbox{.}(2002)]%
        {oelze2002measurement}
\bibfield{author}{\bibinfo{person}{Michael~L Oelze}, \bibinfo{person}{William~D O'Brien}, {and} \bibinfo{person}{Robert~G Darmody}.} \bibinfo{year}{2002}\natexlab{}.
\newblock \showarticletitle{Measurement of attenuation and speed of sound in soils}.
\newblock \bibinfo{journal}{\emph{Soil Science Society of America Journal}} \bibinfo{volume}{66}, \bibinfo{number}{3} (\bibinfo{year}{2002}), \bibinfo{pages}{788--796}.
\newblock


\bibitem[Ojo et~al\mbox{.}(2015)]%
        {ojo2015calibration}
\bibfield{author}{\bibinfo{person}{E~RoTimi Ojo}, \bibinfo{person}{Paul~R Bullock}, \bibinfo{person}{Jessika L’Heureux}, \bibinfo{person}{Jarrett Powers}, \bibinfo{person}{Heather McNairn}, {and} \bibinfo{person}{Anna Pacheco}.} \bibinfo{year}{2015}\natexlab{}.
\newblock \showarticletitle{Calibration and evaluation of a frequency domain reflectometry sensor for real-time soil moisture monitoring}.
\newblock \bibinfo{journal}{\emph{Vadose Zone Journal}} \bibinfo{volume}{14}, \bibinfo{number}{3} (\bibinfo{year}{2015}), \bibinfo{pages}{vzj2014--08}.
\newblock


\bibitem[Rao and Singh(2011)]%
        {rao2011moisture}
\bibfield{author}{\bibinfo{person}{B~Hanumantha Rao} {and} \bibinfo{person}{DN Singh}.} \bibinfo{year}{2011}\natexlab{}.
\newblock \showarticletitle{Moisture content determination by TDR and capacitance techniques: a comparative study}.
\newblock \bibinfo{journal}{\emph{Int. J. Earth Sci. Eng}} \bibinfo{volume}{4}, \bibinfo{number}{6} (\bibinfo{year}{2011}), \bibinfo{pages}{132--137}.
\newblock


\bibitem[Rasheed et~al\mbox{.}(2022)]%
        {rasheed2022soil}
\bibfield{author}{\bibinfo{person}{Muhammad~Waseem Rasheed}, \bibinfo{person}{Jialiang Tang}, \bibinfo{person}{Abid Sarwar}, \bibinfo{person}{Suraj Shah}, \bibinfo{person}{Naeem Saddique}, \bibinfo{person}{Muhammad~Usman Khan}, \bibinfo{person}{Muhammad Imran~Khan}, \bibinfo{person}{Shah Nawaz}, \bibinfo{person}{Redmond~R Shamshiri}, \bibinfo{person}{Marjan Aziz}, {et~al\mbox{.}}} \bibinfo{year}{2022}\natexlab{}.
\newblock \showarticletitle{Soil moisture measuring techniques and factors affecting the moisture dynamics: A comprehensive review}.
\newblock \bibinfo{journal}{\emph{Sustainability}} \bibinfo{volume}{14}, \bibinfo{number}{18} (\bibinfo{year}{2022}), \bibinfo{pages}{11538}.
\newblock


\bibitem[Ravindran and Gratchev(2022)]%
        {ravindran2022effect}
\bibfield{author}{\bibinfo{person}{Sinnappoo Ravindran} {and} \bibinfo{person}{Ivan Gratchev}.} \bibinfo{year}{2022}\natexlab{}.
\newblock \showarticletitle{Effect of water content on apparent cohesion of soils from landslide sites}.
\newblock \bibinfo{journal}{\emph{Geotechnics}} \bibinfo{volume}{2}, \bibinfo{number}{2} (\bibinfo{year}{2022}), \bibinfo{pages}{385--394}.
\newblock


\bibitem[Sakti et~al\mbox{.}(2018)]%
        {sakti2018estimating}
\bibfield{author}{\bibinfo{person}{MBG Sakti}, \bibinfo{person}{DP Ariyanto}, {et~al\mbox{.}}} \bibinfo{year}{2018}\natexlab{}.
\newblock \showarticletitle{Estimating soil moisture content using red-green-blue imagery from digital camera}. In \bibinfo{booktitle}{\emph{IOP Conference Series: Earth and Environmental Science}}, Vol.~\bibinfo{volume}{200}. IOP Publishing, \bibinfo{pages}{012004}.
\newblock


\bibitem[Song et~al\mbox{.}(2024)]%
        {song2024regional}
\bibfield{author}{\bibinfo{person}{Kangle Song}, \bibinfo{person}{Jing Nie}, \bibinfo{person}{Yang Li}, \bibinfo{person}{Jingbin Li}, \bibinfo{person}{Pengxiang Song}, {and} \bibinfo{person}{Sezai Ercisli}.} \bibinfo{year}{2024}\natexlab{}.
\newblock \showarticletitle{Regional soil water content monitoring based on time-frequency spectrogram of low-frequency swept acoustic signal}.
\newblock \bibinfo{journal}{\emph{Geoderma}}  \bibinfo{volume}{441} (\bibinfo{year}{2024}), \bibinfo{pages}{116765}.
\newblock


\bibitem[Steelman and Endres(2011)]%
        {steelman2011comparison}
\bibfield{author}{\bibinfo{person}{Colby~M Steelman} {and} \bibinfo{person}{Anthony~L Endres}.} \bibinfo{year}{2011}\natexlab{}.
\newblock \showarticletitle{Comparison of petrophysical relationships for soil moisture estimation using GPR ground waves}.
\newblock \bibinfo{journal}{\emph{Vadose Zone Journal}} \bibinfo{volume}{10}, \bibinfo{number}{1} (\bibinfo{year}{2011}), \bibinfo{pages}{270--285}.
\newblock


\bibitem[Sun et~al\mbox{.}(2022)]%
        {sun2022aim}
\bibfield{author}{\bibinfo{person}{Yimiao Sun}, \bibinfo{person}{Weiguo Wang}, \bibinfo{person}{Luca Mottola}, \bibinfo{person}{Ruijin Wang}, {and} \bibinfo{person}{Yuan He}.} \bibinfo{year}{2022}\natexlab{}.
\newblock \showarticletitle{Aim: Acoustic inertial measurement for indoor drone localization and tracking}. In \bibinfo{booktitle}{\emph{Proceedings of the 20th ACM Conference on Embedded Networked Sensor Systems}}. \bibinfo{pages}{476--488}.
\newblock


\bibitem[Taneja et~al\mbox{.}(2022)]%
        {taneja2022predicting}
\bibfield{author}{\bibinfo{person}{Perry Taneja}, \bibinfo{person}{Hiteshkumar~Bhogilal Vasava}, \bibinfo{person}{Solmaz Fathololoumi}, \bibinfo{person}{Prasad Daggupati}, {and} \bibinfo{person}{Asim Biswas}.} \bibinfo{year}{2022}\natexlab{}.
\newblock \showarticletitle{Predicting soil organic matter and soil moisture content from digital camera images: Comparison of regression and machine learning approaches}.
\newblock \bibinfo{journal}{\emph{Canadian Journal of Soil Science}} \bibinfo{volume}{102}, \bibinfo{number}{03} (\bibinfo{year}{2022}), \bibinfo{pages}{767--784}.
\newblock


\bibitem[Topp et~al\mbox{.}(1980)]%
        {topp1980electromagnetic}
\bibfield{author}{\bibinfo{person}{G~Clarke Topp}, \bibinfo{person}{J~Lee Davis}, {and} \bibinfo{person}{A~Peter Annan}.} \bibinfo{year}{1980}\natexlab{}.
\newblock \showarticletitle{Electromagnetic determination of soil water content: Measurements in coaxial transmission lines}.
\newblock \bibinfo{journal}{\emph{Water resources research}} \bibinfo{volume}{16}, \bibinfo{number}{3} (\bibinfo{year}{1980}), \bibinfo{pages}{574--582}.
\newblock


\bibitem[Vold{\'a}n et~al\mbox{.}(2024)]%
        {voldan2024moisture}
\bibfield{author}{\bibinfo{person}{Michal Vold{\'a}n}, \bibinfo{person}{Libor Husn{\'\i}k}, {and} \bibinfo{person}{David Mahovsk{\`y}}.} \bibinfo{year}{2024}\natexlab{}.
\newblock \showarticletitle{Moisture estimation by measurement of attenuation of the acoustic resonance mode}.
\newblock \bibinfo{journal}{\emph{Applied Acoustics}}  \bibinfo{volume}{217} (\bibinfo{year}{2024}), \bibinfo{pages}{109847}.
\newblock


\bibitem[Wang et~al\mbox{.}(2020)]%
        {wang2020soil}
\bibfield{author}{\bibinfo{person}{Ju Wang}, \bibinfo{person}{Liqiong Chang}, \bibinfo{person}{Shourya Aggarwal}, \bibinfo{person}{Omid Abari}, {and} \bibinfo{person}{Srinivasan Keshav}.} \bibinfo{year}{2020}\natexlab{}.
\newblock \showarticletitle{Soil moisture sensing with commodity RFID systems}. In \bibinfo{booktitle}{\emph{Proceedings of the 18th International Conference on Mobile Systems, Applications, and Services}}. \bibinfo{pages}{273--285}.
\newblock


\end{thebibliography}

\end{document}